%Paper: hep-th/9206081
%From: NINOMIYA <NINOMIYA%JPNYITP.BITNET@pucc.princeton.edu>
%Date: Mon, 22 Jun 92 19:16:57 JST
%Date (revised): Thu, 25 Jun 92 19:49:13 JST

% (revised) figures added to the 1st version.
% Use plain TeX with phyzzx macro.
% There is a figure appended as a PostScript file at the end of this file.
%
\message{!!!!!!!!!!!!!!!!!!!!!!!!!!!!!!!!!!!!!!!!!!!!!!!!!!!!!!!!!!!!!!!%
!!!!!!!!!}
\message{There is a figure appended as a PostScript file at the end of
this file.}
\message{!!!!!!!!!!!!!!!!!!!!!!!!!!!!!!!!!!!!!!!!!!!!!!!!!!!!!!!!!!!!!!!%
!!!!!!!!!}
\input phyzzx
%%%%% YITP, Uji macros
\def\unlock{\catcode`@=11} % This allows us to modify PLAIN macros.
\def\lock{\catcode`@=12} % at signs are no longer letters
\unlock
%%%%%%%%%%%%%%%%%%%%%%%%%%%%%%%%%%%%%%%%%%%%%%%%%%%%%%%
%% First comes the CORRECTIONS of PHYZZX(YU).TEX macro
%%%%%%%%%%%%%%%%%%%%%%%%%%%%%%%%%%%%%%%%%%%%%%%%%%%%%%
\paperfootline={\hss\iffrontpage\else\ifp@genum\tenrm
    -- \folio\ --\hss\fi\fi}
\def\titlestyle#1{\par\begingroup \titleparagraphs
     \iftwelv@\fourteenpoint\fourteenbf\else\twelvepoint\twelvebf\fi
   \noindent #1\par\endgroup }
\def\GENITEM#1;#2{\par \hangafter=0 \hangindent=#1
    \Textindent{#2}\ignorespaces}
\def\papersize{\hsize=35.2pc \vsize=52.7pc \hoffset=0.5pc \voffset=0.8pc
   \advance\hoffset by\HOFFSET \advance\voffset by\VOFFSET
   \pagebottomfiller=0pc
   \skip\footins=\bigskipamount \normalspace }
\papers  %  This is the default
\def\lettersize{\hsize=35.2pc \vsize=50.0pc \hoffset=0.5pc %
       \voffset=2.5pc
   \advance\hoffset by\HOFFSET \advance\voffset by\VOFFSET
   \pagebottomfiller=\letterbottomskip
   \skip\footins=\smallskipamount \multiply\skip\footins by 3
   \singlespace }
\def\title#1{\vskip\frontpageskip\vfill
   {\fourteenbf\titlestyle{#1}}\vskip\headskip\vfill }
\def\address#1{\par\kern 5pt \titlestyle{\twelvepoint\sl #1}}
\def\abstract#1{\vfill\vskip\frontpageskip\centerline%
               {\fourteencp Abstract}\vskip\headskip#1\endpage}
\newif\ifYITP \YITPtrue
\font\fourteenmib =cmmib10 scaled\magstep2    \skewchar\fourteenmib='177
\font\elevenmib   =cmmib10 scaled\magstephalf   \skewchar\elevenmib='177
\def\YITPmark{\hbox{\fourteenmib YITP\hskip0.2cm
        \elevenmib Uji\hskip0.15cm Research\hskip0.15cm Center\hfill}}
\def\titlepage{\FRONTPAGE\papers\ifPhysRev\PH@SR@V\fi
    \ifYITP\null\vskip-1.70cm\YITPmark\vskip0.6cm\fi %this line is added
   \ifp@bblock\p@bblock \else\hrule height\z@ \rel@x \fi }
%%%%%%%%%%%%%%%%%%%%%%%%%%%%%%%%%%%%%%%%%%%%%%%%
%% end of the CORRECTIONS of PHYZZX(YU).TEX  %%%
%%%%%%%%%%%%%%%%%%%%%%%%%%%%%%%%%%%%%%%%%%%%%%%%

%%%%%%%%%%%%%%%%%%%%%%%
\def\schapter#1{\par \penalty-300 \vskip\chapterskip
   \spacecheck\chapterminspace
   \chapterreset \titlestyle{\ifcn@@\S\ \chapterlabel.~\fi #1}
   \nobreak\vskip\headskip \penalty 30000
   {\pr@tect\wlog{\string\chapter\space \chapterlabel}} }

%%%%%%%%%%%%%%%%%%%%%%%%%
\def\ssection#1{\par \ifnum\lastpenalty=30000\else
   \penalty-200\vskip\sectionskip \spacecheck\sectionminspace\fi
   \gl@bal\advance\sectionnumber by 1
   {\pr@tect
   \xdef\sectionlabel{\ifcn@@ \chapterlabel.\fi
       \the\sectionstyle{\the\sectionnumber}}%
   \wlog{\string\section\space \sectionlabel}}%
   \noindent {\S \caps\thinspace\sectionlabel.~~#1}\par
   \nobreak\vskip\headskip \penalty 30000 }
%%%%%%%%%%%%%%%%%%%%%%%%%%%%%%%%%%%%%%%%%%%%%%%%%%%%%%
%%%%  YITP Uji Res. Ctr. preprint. header part is above
%%%%%%%%%%%%%%%%%%%%%%%%%%%%%%%%%%%%%%%%%%%%%%%%%%%%%%

%
\newtoks\pubnum
\Pubnum={YITP/U-\the\pubnum}
\pubnum={9?-??}
%%%%%%%%%%%%%%%%%%%%%%%%%%%%%%%%%%%%%%%
%%%%  YITP Uji letter head
%%%%%%%%%%%%%%%%%%%%%%%%%%%%%%%%%%%%%%
% fonts for YITP letter head
\font\fourteencmssbx=cmssbx10 scaled\magstep2
\font\twelvecmssbx=cmssbx10 scaled\magstep1
\font\twelvecmss  =cmss12
\font\eightcmss   =cmss8
\def\YITPHEAD{\null\vskip -2.0cm
 \centerline{\fourteencmssbx UJI RESEARCH CENTER}
 \centerline{\twelvecmssbx YUKAWA INSTITUTE FOR THEORETICAL PHYSICS}
 \vskip-0.0cm\centerline{\twelvecmss Kyoto University, Uji 611, Japan}
  {\eightcmss\baselineskip=0.30cm\tabskip=0pt plus400pt
  \halign to \hsize{\hskip11.3cm ## & ## & ## & ## \cr
  \ &\hskip1pt PHONE    &:&0774--20--7421\cr
  \ &\hskip1pt FAX      &:&0774--33--6226\cr}}
  \vskip 7mm}
\def\YITPletter{\lettersize \let\headline=\letterheadline
        \let\footline=\letterfootline \FRONTPAGE\YITPHEAD\addressee}
%%%%%%%%%%%%%%%%%%%%%%%%%%%%%%%%%%%%%%

%

%

%
\def\globaleqnumbers{\relax\if\equanumber<0\else\global\equanumber=-1\fi}
%
%%%%% Add equation number by 1
\def\addeqno{\ifnum\equanumber<0 \global\advance\equanumber by -1
    \else \global\advance\equanumber by 1\fi}
%%%%%%%%%%%%%%%%%%%%%%%%%%%
\mathchardef\Lag="724C  %%%% Lagrangian symbol = \Lag

 %%  definition of \yen mark
%
%%End equation mark with equation number.
 \def\ee{\eqno\eq }
%%Draw a box around a thing.
 
%%Figure mark in text
 
%%Table mark in text
 
%%Over Bar ---> Use \overline
 \def\overbar#1{\vbox{\ialign{##\crcr
           \vrule depth 2mm
           \hrulefill\vrule depth 2mm
           \crcr\noalign{\kern-1pt\vskip0.125cm\nointerlineskip}
           $\hfil\displaystyle{#1}\hfil$\crcr}}}
%% Enforce carriage return and line feed within a paragraph

%%rho with a little bit up

%
 %Equation number as you like
%
%A little bigger cdot

%%%%%%%%%%%
%square (in math mode)
%
\def\sqr#1#2{{\vcenter{\hrule height.#2pt
      \hbox{\vrule width.#2pt height#1pt \kern#1pt
          \vrule width.#2pt}
      \hrule height.#2pt}}}

%%%%%%%%%

\def\Buildrel#1\under#2{\mathrel{\mathop{#2}\limits_{#1}}}
\def\llongrarrow{\hbox to 40pt{\rightarrowfill}}

%%%%%%%%%%
\def\journals#1&#2(#3){\unskip; \sl #1~\bf #2 \rm (19#3) }
%%%%%%%%%%%%%%%%%%%%%%%%%%%%%%%%%%%
%% item number at the leftmost end
%
 \def\nllap#1{\hbox to-0.35em{\hskip-\hangindent#1\hss}}

%%%%%%%%%%%%%%%%%%%%%%

%
\def\rslash{\partial\kern-0.026em\raise0.17ex\llap{/}%
          \kern0.026em\relax}
\def\Dslash{D\kern-0.15em\raise0.17ex\llap{/}\kern0.15em\relax}
%%%%%%%%%%%%%%%%%%%%%%%
\mathchardef\bigtilde="0365

%%%%%%%%%%%%%%%%%%%%%%%%%%%%%%%%%%%%%%%%
%   double eqalign macro (\deqalign{...})
%
\def\deqalign#1{\null\,\vcenter{\openup1\jot \m@th
    \ialign{\strut\hfil$\displaystyle{##}$&$\displaystyle{##}$&$
	\displaystyle{{}##}$\hfil\crcr#1\crcr}}\,}
%%%%%%%%%%%%%%%%%%%%%%%%%%%%%%%%%%%%%%%%%%%%%%%%%%%%%%%%%%
%
% [ \seqalignno{..&..&\seq\cr} \eqs ]
%
%  "\seqalignno" gives alignned equations with eq-numbers like
%     (1.2a),(1.2b) ... with "\seqn{...}" and/or "\seq"[=\seqn\?].
%  "\eqsname{...}" or "\eqs"[=\eqsname\?] gives an eq-number
%     like (1.2) to the alignned equations (1.2a),(1.2b)... .
%     ("\seqn" can be used also in \eqalignno: \seqn=\eqnalign.)
%
\newcount\eqabcno \eqabcno=97 %% \char97=``a''
\newcount\sequanumber \sequanumber=0
\newif\iffirstseq \firstseqtrue
\newif\ifequationsabc \equationsabcfalse
\def\eqnameabc#1{\relax\pr@tect
    \iffirstseq\global\firstseqfalse%
        \ifnum\equanumber<0 \global\sequanumber=\number-\equanumber
           \xdef#1{{\rm(\number-\equanumber a)}}%
           \global\advance\equanumber by -1
        \else \global\advance\equanumber by 1
           \xdef#1{{\rm(\ifcn@@ \chapterlabel.\fi \number\equanumber a)}}
        \fi
    \else\global\advance\eqabcno by 1
        \ifnum\equanumber<0
           \def#1{{\rm(\number\sequanumber \char\number\eqabcno)}}%
        \else
           \xdef#1{{\rm(\ifcn@@ \chapterlabel.\fi%
                        \number\equanumber \char\number\eqabcno)}}
        \fi
    \fi}

\def\eqsname#1{\relax\pr@tect%
        \ifnum\equanumber<0
           \def#1{{\rm(\number\sequanumber)}}
        \else
           \xdef#1{{\rm(\ifcn@@ \chapterlabel.\fi \number\equanumber)}}
        \fi}

%%%%%%%%%%%%%%%%%%%%%%%%%%%%%%%%%%%%%%%%%%%%%%%%%%%%%%%%%%%%%%%%%%%%%%%
%% \draft causes the symbolic names of equations to be printed
%% alongside the equation numbers

%%%%%%%%%%%%%%%%%%%%%%%%
\let\Title=\title  % to save previous macro.
\lock
%%%%%%%%%%%%%%%%%%%%%%%%%%%%%%%%%%%%%%%%%%%%%%%%%%%%%%%%%%%%%%%%%%%%%%%
% Here begins the text.
\Pubnum={\the\pubnum}
\pubnum={UT-614\cr TIT/HEP-191\cr YITP/U-92-05\cr hep-th@xxx/9206081}
\date={June 1992}
\titlepage
\Title{\seventeenbf Scaling Exponents in Quantum Gravity near Two
Dimensions}
\author{
Hikaru Kawai \footnote*{{\twelvesl Department of Physics, University of
Tokyo, Hongo, Tokyo 113, Japan}
\quad\quad\quad E-mail address : TKYVAX\$hepnet::KAWAI},
Yoshihisa Kitazawa \footnote\dag{{\twelvesl Department of Physics, Tokyo
Institute of Technology, Oh-okayama, Meguro-ku, Tokyo 152, Japan }
\hfil\break E-mail address : TITVS0::KITAZAWA.decnet},
and Masao Ninomiya \footnote\#{{\twelvesl
Uji Research Center, Yukawa Institute for Theoretical Physics, Kyoto
University, Uji 611, Japan }
\hfil\break E-mail address : NINOMIYA@JPNYITP.bitnet}
}
\def\abstract#1{\vfill\vskip\frontpageskip\centerline%
               {\fourteenbf Abstract}\vskip\headskip#1\endpage}
\abstract{
We formulate quantum gravity in $2+\epsilon$ dimensions in such a way
that the conformal mode is explicitly separated.
The dynamics of the conformal mode is understood in terms of the
oversubtraction due to the one loop counter term.
The renormalization of the gravitational dressed operators is studied
and their anomalous dimensions are computed.
The exact scaling exponents of the 2 dimensional quantum gravity are
reproduced in the strong coupling regime when we take
$\epsilon\rightarrow0$ limit.
The theory possesses the ultraviolet fixed point as long as the
central charge $c<25$, which separates weak and strong coupling phases.
The weak coupling phase may represent the same universality class with
our Universe in the sense that it contains massless gravitons if we
extrapolate $\epsilon$ up to 2.
}
%%%%%%%%%%%%%%%%%%%%%%%%%%%%%%%%%%%%%%%%%%%%%%%%%%%%%%%%%%%%%%%%%%%%%
\chapter{Introduction}
Our attempt to understand quantum gravity is plagued by multitudes of
difficulties such as non-renormalizability and the instability of the
conformal mode. In order to circumvent these problems, various
approaches to make sense of quantum gravity have been proposed with
considerable success. We may cite string theory and topological
gravity as examples. Of all these attempts, notable progress in our
understanding of quantum gravity is brought about by the exact
solution of two dimensional quantum gravity. Although it has been a
toy model, this has provided us precious insight into more general
universal classes of quantum gravity.

In view of the existence of two dimensional quantum theory of
gravitation, it is reasonable to expect that consistent quantum
gravity theory exists in 2+$\epsilon$ dimensions as well.
In this context one may draw an analogy with nonlinear sigma
models. Nonlinear sigma models are renormalizable in two dimensions
and they are asymptotically free.
Furthermore they possess well defined 2+$\epsilon$ dimensional
expansion. Such an expansion provides us with information about
realistic nonlinear sigma models which are relevant to critical
phenomena in Nature.

In the constructive (lattice) approach to quantum gravity, dynamical
triangulation method has turned out to be effective to integrate over
the metric. This point has been demonstrated by the success of the
matrix model approach to the 2 dimensional quantum gravity.
The dynamical triangulation method has
been extended to higher dimensions such as 3 and 4 dimensional quantum
gravity [1, 2]. They have indicated the existence of two distinct
phases in 3 and 4 dimensional quantum gravity. This feature appears to
be in accord with the $2 + \epsilon$ dimensional expansion approach
[3 - 6].
We believe it is important to develop this analytical approach further.

The organization of this paper is as follows.
In section 2, we formulate quantum gravity in $2+\epsilon$ dimensions
in such a way that the conformal mode is explicitly separated.
In section 3, the one loop counter term is computed in our formulation.
The dynamics of the conformal mode is understood in terms of the
oversubtraction due to the one loop counter term.
In section 4, the renormalization of the gravitational dressed
operators is studied and their anomalous dimensions are computed.
The two dimensional quantum gravity turns out to be
$\epsilon\rightarrow0$ limit of the strong coupling regime.
The section 5 is devoted to the conclusions and discussions.
We explain the basic physical picture of the $2+\epsilon$ dimensional
quantum gravity.
The appendix contains the short summary of the background field
method in our formulation.

\chapter{Separation of conformal mode in $2+\epsilon$ dimensional
gravity}

Firstly we would like to seek a proper formalism of $D = 2 + \epsilon$
dimensional quantum gravity. {}From the study of two dimensional
quantum gravity, we have learned that the conformal mode of the
metric is the important dynamical degree of freedom in two dimensions.
Therefore our strategy is to adopt a parametrization and a gauge which
singles out the conformal mode. Let us write the metric as follows
$$
 g_{\mu\nu} = {\hat g}_{\mu\rho} {(e^h)}^\rho\,_\nu ~ e^{-\phi}
  \ee
$$
where ${\hat g}_{\mu\nu}$ is the background metric and the
${h^\mu}_\nu$ field is taken to be traceless ${h^\mu}_\mu = 0$.
Hence
$$
 \sqrt{g} = \sqrt{{\hat g}} ~ e^{-{D\over2}\phi} ~.   \ee
$$
We define further
$$
 \eqalign{ {\tilde g}_{\mu\nu} & = {\hat g}_{\mu\rho}
             {{(e^h)}^\rho}_\nu \cr
           & = {\hat g}_{\mu\nu} + h_{\mu\nu} + {1\over2}
     h_{\mu\rho} {\hat g}^{\rho\sigma} h_{\sigma\nu} + \cdots \cr}
\ee
$$
where tensor indices are raised and lowered by the background metric
${\hat g}_{\mu\nu}$ and $h_{\mu\nu}$ is symmetric in $\mu$ and $\nu$.

The Einstein action becomes in terms of these variables
$$
\eqalign{
 \int d^D x \sqrt{g} R  & =
    \int d^D x \sqrt{\hat g} ~ e^{-{\epsilon\over2}\phi} {\tilde R}
\cr
    & - \int d^D x \sqrt{{\hat g}} ~ e^{-{\epsilon\over2}\phi} ~
   {1\over4} \epsilon (D-1) {\tilde g}^{\mu\nu} \partial_\mu \phi
   \partial_\nu \phi  ~ .
\cr
	} \ee
$$
We expand the action in terms of ${h^\mu}_\nu$ and $\phi$ fields and
drop the linear term in these fields
$$
\eqalign{
 \int d^D x \sqrt{g} R
       & = \int d^D x \sqrt{{\hat g}}{\hat R}
\cr
       & + \int d^D x \sqrt{{\hat g}} ~ \lbrace
         {1\over 4}{h^\mu}_{\nu,\rho} {h^\nu}_\mu,^\rho
  +{1\over 2}\hat R^\sigma\,_{\mu\nu\rho}{h^\rho}_\sigma{h^{\mu\nu}}
\cr
       &-{\epsilon\over 4}(D-1){{\hat g}}^{\mu\nu}
	\partial_\mu \phi \partial_\nu \phi
          + {{\epsilon^2}\over 8} \phi^2 {\hat R}
          +{\epsilon\over 2} \phi {h^\mu}_\nu {\hat R}^\nu\,_\mu
\cr
       &+{\epsilon\over 2}\phi h^{\mu\nu}, _{\mu\nu}
        -{1\over 2}{h^\nu}_{\mu , \nu} h^{\rho \mu},_{\rho}\rbrace
        + \cdots ~.
\cr
	}
\ee
$$
In this paper we use the notation of t'Hooft and Veltman [7].
In particular ,$\mu$ denotes the covariant derivative with respect
to the background metric.

When we compute the one loop counter term, we use the background
field method. This is an efficient and manifestly gauge invariant
method to compute the effective action. We drop the term linear
in quantum fields in the action, because it is implicitly assumed
that such fields are coupled to sources which drive them to assume
their background form.
Therefore if we use the conventional coupling of the source to the
metric $g_{\mu\nu}J^{\mu\nu}$, we should have dropped the
linear term in ${h'}_{\mu\nu}$, where
$g_{\mu\nu} = {\hat g}_{\mu\nu} + {h'}_{\mu\nu}$,
rather than ${h^\mu}_\nu$ and $\phi$ fields separately.
However we have
checked that the difference does not lead to any change in the one
loop counter term.
See the appendix for more detailed discussions.

In order to cancel the last two terms in the quadratic action, we
choose the following gauge fixing term
$$
\int d^D x {\sqrt{\hat g}} ~ {1\over2}
  ({h^\nu}_{\mu,\nu} + {\epsilon\over2}\partial_\mu \phi)
  ({h^{\rho \mu}}_{,\rho} + {\epsilon\over2}\partial^\mu \phi) ~.
  \ee
$$
Including this term, the total quadratic action is
$$
\eqalign{
 \int d^D x {\sqrt{\hat g}} ~
	&  \lbrace
	  {1\over 4}{h^\rho}_{\mu , \nu} {{h^\mu}_{\rho ,}}^\nu
   +{1\over 2}{\hat R}^\sigma\,_{\mu\nu\rho} {h^\rho}_\sigma
		h^{\mu\nu}
\cr
	& -{\epsilon\over 8}D {\hat g}^{\mu\nu}
		\partial_\mu \phi \partial_\nu \phi
	+ {\epsilon\over 2}\phi {h^\mu}_\nu {\hat R}^\nu\,_\mu
	+ {{\epsilon^2}\over 8} \phi^2 {\hat R}
					\rbrace ~.
\cr}
\ee
$$
For later convenience we write down some of the interaction
vertices which are readable from eq.(2.4)
$$
\eqalign{ \int d^D x \sqrt{{\hat g}} ~ & \lbrace
{1\over 8} \epsilon^2
(D-1) \phi {\hat g}^{\mu\nu} \partial_\mu \phi \partial_\nu
\phi - {1\over 32}\epsilon^3 (D-1) \phi^2 {\hat g}^{\mu\nu}
\partial_\mu \phi \partial_\nu \phi \cr
& + {1\over 4} \epsilon (D-1) h^{\mu\nu} \partial_\mu \phi
\partial_\nu \phi - {1\over 8} \epsilon (D-1) {h^\mu}_\rho
h^{\rho\nu} \partial_\mu \phi \partial_\nu \phi \cr
& + \cdots  \rbrace ~.  \cr}
\ee
$$
Under the general coordinate transformation,
$$
g_{\mu\nu} \rightarrow g_{\mu\nu} + \partial_\mu \epsilon^\rho
g_{\rho\nu} + g_{\mu\rho} \partial_\nu \epsilon^\rho +
\epsilon^\rho \partial_\rho g_{\mu\nu}
\ee
$$
${h^\mu}_\nu$ and $\phi$ fields transform as follows :
$$
\eqalign{
 \delta h^\mu\,_\nu
    & = {(e^{-h})^\mu}_\rho
	{\hat g}^{\rho\sigma} {\epsilon^\tau}_{, \sigma}
	{({\hat g} e^h)}_{\tau\nu} + \epsilon^\mu, _{\nu}
\cr
    &\quad +{{(e^{-h})}^\mu}_\rho {{(e^h)}^\rho}_{\nu ,\sigma}
	\epsilon^\sigma
	- {2\over D} \epsilon^\rho, _\rho {\delta^\mu}_\nu
\cr
    & \quad - {1\over D} {{(e^{-h})}^\rho}_\sigma
	{{(e^h)}^\sigma}_{\rho , \tau}
	\epsilon^\tau \delta^\mu\,_\nu
\cr
\delta \phi & = \epsilon^\mu \partial_\mu \phi
	- {2\over D}\epsilon^\mu , _\mu
	- {1\over D}{{(e^{-h})}^\mu}_\nu
		{{(e^h)}^\nu}_{\mu , \rho} \epsilon^\rho ~.
\cr
	}   \ee
$$
Under these transformations, the gauge fixing term changes as
$$
\eqalign{
 \delta & ({h^\nu}_{\mu , \nu}+{\epsilon\over 2}\partial_\mu \phi)
\cr
& = {\epsilon_\mu},^\nu\,_\nu - {\hat R}_{\mu\nu} \epsilon^\nu
    + {\epsilon\over 2} (\partial_\nu \phi \epsilon^\nu)_{, \mu}
     + \cdots  ~ .
\cr}  \ee
$$
Therefore the ghost action is
$$
 \sqrt{\hat g} \lbrace{
	\psi^{\ast\mu} {\psi_{\mu ,}}^\nu\,_\nu
	  - \psi^{\ast\mu}{\hat R}^\nu\,_\mu \psi_\nu
		- {\epsilon\over 2}(\partial^\nu \phi)
			\psi^{\ast \mu},_\mu \psi_\nu + \cdots
		       }\rbrace
\ee
$$
where only the coupling of the $\phi$ field to the ghost field is
shown and it is proportional to $\epsilon$.

$h_{\mu\nu}$ is a symmetric matrix and $\hat g^{\mu\nu}h_{\mu\nu}=0$.
We introduce a traceless symmetric matrix $H_{\mu\nu}$
$(\delta^{\mu\nu}H_{\mu\nu}=0)$ and express $h_{\mu\nu}$ as
$$
 {h^\mu}_\nu = {\hat g}^{\mu\rho} H_{\rho\nu}
	-{1\over D}{\delta^\mu}_\nu {\hat g}^{\rho\sigma}
H_{\rho\sigma}\quad .
\ee
$$
Then
$$
{\sqrt{\hat g}} h_{\mu\nu , \rho} {{h^{\mu\nu}},}^ \rho =
\sqrt{\hat g} ({\delta^\mu}_{\mu '} {\delta^\nu}_{\nu '}
- {1\over D}
 {\hat g}^{\mu\nu}{\hat g}_{\mu ' \nu '}) H_{\mu\nu , \rho}
H^{\mu ' \nu '},^ \rho  \ee
$$
where indices are raised by ${\hat g}^{\mu\nu}$ as usual.

In order to proceed further, we expand the background metric around
the flat metric
$$
 {\hat g}_{\mu\nu} = \delta_{\mu\nu} + {\hat h}_{\mu\nu} \quad .
\ee
$$
The kinetic terms of ${h^\mu}_\nu$, $\psi_\mu$ and $\phi$ fields can
be expanded in ${\hat h}_{\mu\nu}$ as well,
$$
\eqalign{
 {1\over 4}
	& \sqrt{\hat g} h_{\mu\nu , \rho} h^{\mu \nu},^\rho
\cr
	& = {1\over 4}({\delta^\mu}_\rho {\delta^\nu}_\sigma
	    - {1\over D}\delta^{\mu\nu} \delta_{\rho\sigma})
	(\partial_\alpha H^{\rho\sigma}
	    +{\hat \eta_{\alpha ,}}^{\rho\sigma}\,_{\rho ' \sigma '}
		H^{\rho ' \sigma '} )
\cr
	&\qquad (\partial_\alpha  H_{\mu\nu}
   - {\hat \eta_{\alpha ,}}^{\mu'\nu'}\,_{\mu \nu} H_{\mu ' \nu '})
	  -{1\over 4} {\hat S}_{\mu\nu}
		\partial_\mu H_{\rho\sigma}
		\partial_\nu H^{\rho\sigma} + \cdots
\cr
	}
\ee
$$
where
$$
\eqalign{
 \hat \eta_\alpha ,^{\rho ' \sigma '}\,_{\rho\sigma}
	 & ={{\hat \Gamma}_\alpha}\,^{\rho '}\,_\rho
		{\delta^{\sigma'}}\,_\sigma
           +{{\hat \Gamma}_\alpha}\,^{\rho '}\,_\sigma
		{\delta^{\sigma'}}_{\rho}
\cr
 {\hat S}_{\alpha\beta}
    & = {\hat h}_{\alpha\beta} - {1\over D}
		\delta_{\alpha\beta} {\hat h}_{\mu\mu}
\cr
	}
\ee
$$
and for $\psi^\mu$
$$
\eqalign{
  \sqrt{\hat g}
   & {\hat g}^{\alpha\beta}\psi^\ast_\mu , _\alpha \psi^\mu , _\beta
\cr
   & = (\partial_\alpha \psi^\ast_\mu - {\hat \eta_\alpha}^\rho\,_\mu
        \psi^\ast_\rho)
	(\partial_\alpha \psi^\mu + {\hat \eta_\alpha}
           ^\mu\,_\sigma \psi^\sigma)
\cr
   & -{\hat S}_{\alpha\beta} \partial_\alpha
               \psi^\ast_\mu \partial_\beta \psi^\mu + \cdots ~ .
\cr
	}
\ee
$$
with
$$
 {{\hat \eta_\alpha}\,^\rho}\,_\mu =
{{{\hat \Gamma}_\alpha}\,^\rho}\,_\mu .
\ee
$$
For $\phi$
$$
  {D\over 8}\epsilon \sqrt{\hat g} ~ {\hat g}^{\alpha\beta}
\partial_\alpha \phi \partial_\beta \phi
 = {D \over 8} \epsilon \partial_\alpha \phi \partial_\alpha \phi
- {D \over 8} \epsilon {\hat S}_{\alpha\beta} \partial_\alpha \phi
 \partial_\beta \phi + \cdots  ~ .  \ee
$$
The propagators are
$$
 \eqalign{ <H_{\mu\nu}(P) H_{\rho\sigma} (-P)> & =
{1\over P^2}
(\delta_{\mu\rho} \delta_{\nu\sigma} + \delta_{\mu\sigma}
  \delta_{\nu\rho}-{2\over D} \delta_{\mu\nu}
   \delta_{\rho\sigma}) \cr
<\psi^\mu (P) \psi^\ast_\nu (-P)> & = {1\over P^2}
{\delta^\mu}_\nu \cr
<\phi(P) \phi(-P)> & = -{1\over P^2}{4\over {\epsilon D}} \quad .
                    \cr } \ee
$$

As is seen, the propagator of the $\phi$ field has a $1\over\epsilon$
singularity. However this singularity is likely to be controllable
since the coupling of the $\phi$ field are suppressed by powers of
$\epsilon$. We demonstrate this point later by concrete calculations.

\chapter{Dynamics of conformal mode at one loop level}

We are now in the position to evaluate the one loop divergences.
It is found that there is no $\eta_\mu$ dependence
in the counter term. To verify this, it is convenient to adopt
the doubling trick and
complexify the $H_{\mu\nu}$ field. Then the same argument goes through
with [4]. ${h^\mu}_\nu$ field gives rise to the following tadpole
divergence
$$
  \sqrt{\hat g} ~ {1\over 2} {{\hat R}^\sigma}\,_{\mu\nu\rho}
<{h^\rho}_\sigma h^{\mu\nu}>
 = \sqrt{\hat g} ~ {1\over{2\pi}}{{\hat R}\over \epsilon} \quad .
 \ee
$$
A tadpole divergence also arises due to the ghost field
$$
 \sqrt{\hat g} {{\hat R}^\nu}\,_\alpha <\psi^{\ast\alpha} \psi_\nu>
= \sqrt{\hat g} {1\over{2\pi}} {{\hat R}\over \epsilon}
\quad . \ee
$$
The remaining divergences come from the well known conformal
anomaly due to free fields. One loop divergence due to a free
scalar field is
$$
 -{1\over{24\pi}} \sqrt{\hat g} {{\hat R}\over \epsilon} \quad .
\ee
$$
The remaining labor in determining the one loop counter term is to
count the number of degrees of freedom.
The $\phi$ field contributes one and
the ${h^\mu}_\nu$ field has two dynamical degrees of freedom in two
dimensions. The complex two component ghost field should be counted
as $-4$. In addition we include matter fields with the central
change c.

In this way we have found that the one loop counter term is
$$
-{{25-c}\over{24\pi}} {1\over \epsilon} \sqrt{\hat g}{\hat R}\quad .
\ee
$$
The bare coupling is
$$
{1\over G_0} = \mu^\epsilon ({1\over G}-{{25-c}\over{24\pi}}
{1\over \epsilon}) \ee
$$
where $G$ is the renormalized coupling. The $\beta$ function is
determined by
$$
 \mu {\partial\over {\partial \mu}}({1\over {G_0}}) = 0 \ee
$$
yielding
$$
 \beta = \epsilon G - {{25-c}\over{24\pi}} G^2 ~. \ee
$$
The $\beta$ function shows that quantum gravity in $2 + \epsilon$
dimensions possesses an ultraviolet fixed point
$G^*={24\pi\over 25-c}\epsilon$ as long as $c<25$.

We would like to recall [6] that the $\beta$ function depends on which
interaction is compared with the gravitational interaction.
It is shown later that the coefficient of the Thirring interaction,
which becomes dimensionless in two dimensions, is automatically fixed
to be unity. Therefore our computation scheme has avoided the double
expansion in $\epsilon$ and the central charge c.

%%%%
Let us examine more closely what we have done in the one loop
renormalization of the theory.
We have considered the tree level action
$$
{\mu^\epsilon\over G}\int d^D x
\left\{ \sqrt{g} R + \sqrt{\hat g} {1\over 2\alpha}
(h^\nu_{~~\mu,\nu}+{\epsilon\over2}\partial_\mu \phi)
(h^{\rho\mu}_{~~~,\rho}+{\epsilon\over2}\partial^\mu\phi)
\right\} \ee
$$
where $\alpha$ is put to 1 in our Feynman type gauge.
We have found that the one loop bare action is the same form with
the tree level action with the substitution
$\mu^\epsilon/G \rightarrow 1/G_0$.
With this coupling constant renormalization, we obtain the
following finite effective action at the one loop level
$$
{1\over G}\mu^\epsilon \int d^D x \sqrt{\hat g}\hat R +
{\rm finite~terms}.\ee
$$
The counter term $-\int d^D x {25-c\over24\pi}
{\mu^\epsilon\over\epsilon} \sqrt{g} R$ is appropriate to the
$h_{\mu\nu}$ field but it is in fact an oversubtraction for the
conformal mode $\phi$.
As it can be seen from eq.(2.4), the conformal mode $\phi$ is
suppressed at least by single power of $\epsilon$.
Therefore there is no divergence which involves the conformal mode
$\phi$ at the one loop level.

Nevertheless we are subtracting a finite term for $\phi$, namely
$\int d^D x {25-c\over 24\pi} {1\over4}\partial_\mu\phi
\partial_\mu\phi$ which is contained in the counter term.
It is an oversubtraction since at the tree level, the kinetic term
of the conformal mode is $O(\epsilon)$.

%%%%
It is an important point and we would like to explain it in detail.
Consider the perturbative evaluation of the effective action around
the flat metric.
The background metric can be decomposed into the conformal mode
$\bar\phi$ and the traceless symmetric matrix $\bar h_{\mu\nu}$,
$\hat g_{\mu\nu}=e^{-\bar\phi}(e^{\bar h})_{\mu\nu}$.
The singularity of the two point function of $\bar h_{\mu\nu}$ field
in the one loop effective action is $O({1\over\epsilon})$.
However the two point function of $\bar\phi$ in the effective action
at the one loop level is $O(\epsilon)$ since there is a factor
$\epsilon$ suppression for each $\bar\phi$ field.
This is very strange since the kinetic term of $\bar\phi$ which is
contained in the one loop counter term is $O(1)$.

The resolution of this puzzle must be that in the full effective
action, a finite nonlocal term is present which precisely cancels
the $O(1)$ term which involves the conformal mode.
In fact the Liouville action is such a term,
$$
\eqalign{
&-{25-c \over 96\pi}\mu^\epsilon \int d^D x \sqrt{\hat g}{\hat R}
{1\over\Delta}{\hat R}\cr
&=-{25-c \over 96\pi}\int d^D x e^{-{\epsilon\over2}\bar\phi}
\left({\hat R}{1\over\Delta}{\hat R}-2\bar\phi\hat R+
\bar\phi\Delta\bar\phi +O(\epsilon)\right)\cr
}\ee
$$
After subtracting the one loop local counter term from the effective
action, the nonlocal Liouville term remains in the effective action.

The remarkable point is that the one loop counter term and the
nonlocal Liouville term are the same as long as the conformal mode
is involved at $O(1)$.
In the conventional field theory, the counter term is present to
cancel the divergent part of the effective action.
Therefore the counter term is not a large quantity in spite of the
${1\over\epsilon}$ pole.
The situation here is different.
As long as the conformal mode is concerned, we are subtracting the
$O(1)$ quantity from the $O(\epsilon)$ quantity.
In this sense, the counter term for the conformal mode is a large
quantity and we are performing an oversubtraction.

%%%%
We are forced to adopt such an oversubtraction in order to respect
the general covariance of the theory.
If we consider the multiple insertions of this counter term, it could
cause extra singularities in ${1\over\epsilon}$.

We argue that this problem can be taken care of by using the following
bare coupling in the calculation,
$$
\mu^\epsilon G_0
= {G\over {1-{25-c\over 24\pi}{G\over\epsilon}}}
= G \sum_{n=0}^\infty \left( {25-c\over 24\pi}{G\over\epsilon}
\right)^n\ee
$$
For the propagator of the conformal mode, the use of the above
mentioned bare coupling is nothing but the insertion of the counter
term $\int d^D x {25-c\over 24\pi} {1\over4}\partial_\mu\phi
\partial_\mu\phi$ infinite times.
Since the counter term dominates over the tree term, such
resummation is necessary for the conformal mode.

The gauge parameter $\alpha$ will be renormalized also.
In order to perform such renormalization, the quantum propagator
(versus background) for $h_{\mu\nu}$ should be calculated.
Although we do not need such a knowledge in this paper, it will be
necessary in the two loop renormalization of the theory[8].

In the following discussion concerning the operator renormalization,
we perform calculations in terms of $G_0\mu^\epsilon$.
As it will be demonstrated shortly such a calculation is free from
the oversubtraction problem.
In particular we can consider $\epsilon\rightarrow 0$ limit
( 2 dimensional quantum gravity).

\chapter{Gravitational anomalous dimensions and two dimensional limit}

We next introduce the cosmological constant term $\int d^D x \sqrt{g}$
 into the action (2.4) and evaluate the anomalous dimension
$\gamma_{\Delta_0}$,
where $\Delta_0 = 0$ denotes
the canonical dimension. In the present article it is assumed that
this operator is multiplicatively renormalizable.
In our parametrization (2.1), the cosmological constant operator
takes the form (2.2).
In the following we compute the one and two loop corrections to
$\sqrt{{\hat g}}$ by taking into account only gravitational
fluctuations.

To evaluate the divergent part of each graph, we have made assumption
that $G_0 \mu^\epsilon $ is of order
$\epsilon (= D-2)$.
%%%%
Due to the relation eq.(3.5), it is the case as long as
$G\gsim\epsilon$. At the one loop level we obtain
$$
 <{1\over{2!}} {({D\over 2}\phi)}^2> =
	{{G_0 \mu^\epsilon}\over{2\pi \epsilon}}
	({1\over\epsilon} + \hbox{const.} ) ~. \ee
$$
At the two loop order there are 9 graphs as depicted in Fig.1.
We notice from (2.8) that $\phi$ propagator (2.21) is order
$1\over\epsilon$ and $3\phi$ and $4\phi$ vertices are proportional to
$\epsilon^2$ and $\epsilon^3$ respectively, while the $\phi-2h$ and
$2\phi-2h$
vertices are proportional to $\epsilon$. Using these facts we may
determine the strength of the $1\over\epsilon $ singularity for each
graph. Under the assumption $G_0 \mu^\epsilon \sim 0(\epsilon)$,
graphs (d),(e) and (g) turn out to be finite. The divergent part
of graphs (c) cancels with that of (f).
%%%
The ghost contribution graph (h) is also finite due to the
suppression factor $\epsilon$ in eq.(2.12).
The same is true for the matter contribution depicted by graph (i).
%%%
The remaining divergences
of the two loop graphs consist of the graphs (a) and (b).
The divergence of graph (a) is given by
$$
{1\over 2}({{G_0 \mu^\epsilon}\over{2\pi \epsilon}})^2
{({1\over \epsilon} + \hbox{const.})}^2 ~. \ee
$$
For graph (b) we obtain
$$
-{1\over 2}\,{1\over \epsilon}
	({G_0 \mu^\epsilon \over 2\pi \epsilon})^2.
\ee
$$

Summarizing the calculation up to the two loop order, we obtain the
renormalized cosmological term
$Z \sqrt{{\hat g}} ~ e^{-{D\over2}\phi}$,
where $Z$ is given by
$$
 Z = 1-({{G_0 \mu^\epsilon}\over{2\pi \epsilon}})
     ({1\over \epsilon} + \hbox{const.})
      +{1\over 2}\,{1\over \epsilon}
        ({{G_0 \mu^\epsilon}\over{2\pi \epsilon}})^2
      +{1\over 2}({{G_0 \mu^\epsilon}\over{2\pi \epsilon}})^2
        ({1\over \epsilon} + \hbox{const.})^2 \quad .
\ee
$$
In order to evaluate the anomalous dimension $\gamma_{\Delta_0 = 0}$
defined by
$$
 \gamma_{\Delta_0} = \mu {\partial \over \partial \mu} \log \, Z
\ee
$$
we make an expansion of $ \log Z$ as
$$
 \log Z = - {1\over \epsilon} {{G_0 \mu^\epsilon}\over{2\pi\epsilon}}
	+ {1\over 2}{1\over \epsilon}
	({{G_0 \mu^\epsilon}\over{2\pi \epsilon}})^2 +O(G_0\,^3)
\quad .
\ee
$$
Thus $\gamma_{{\Delta_0} = 0}$ is obtained as
$$
 \gamma_{{\Delta_0} = 0} =
- {{G_0 \mu^\epsilon}\over{2\pi \epsilon}}
+ {({{G_0 \mu^\epsilon}\over{2\pi \epsilon}})}^2 + O ({G_0}^3) \quad .
\ee
$$

The one loop coupling $G_0 \mu^\epsilon$ has already been computed in
eq.(3.5).
However when evaluating $\gamma_{\Delta_0 = 0}$ in eq.(4.7) we assume
that in the bare Lagrangian the counter term is dominant so that $G_0$
is given by
$$
 {1\over{G_0}} = - {{25-c}\over{24\pi}} {{\mu^\epsilon}\over \epsilon}
\ee
$$
This assumption is equivalent to say that the renormalized coupling
constant $G$ is not small $G \gg \epsilon$. In this case we obtain
$$
 \gamma_{{\Delta_0} = 0} = {4\over {Q^2}} + {({4\over {Q^2}})}^2 +
O ({({4\over {Q^2}})}^3)
\ee
$$
where $Q = \sqrt{{25-c}\over3}$ in the notation [9].

For a general operator $\int d^D x {\sqrt{g}}^{1-{\Delta_0}} \Phi$
of a spinless field $\Phi$ with a canonical dimension $2\Delta_0$, we
have performed a similar calculation up to two loops for the operator
${\sqrt{g}}^{1-{\Delta}_0} = {\sqrt{{\hat g}}}^{1-\Delta_0}
e^{-{D\over2}(1 - \Delta_0)\phi}$. The anomalous dimension is
obtained as
$$
\gamma_{\Delta_0} = {4\over {Q^2}}
 (1 - \Delta_0)^2 +
 ({4\over{Q^2}})^2 (1-\Delta_0)^3 + O(({4\over {Q^2}})^3)\quad .
\ee
$$
%%%
Let us recall the fermion field coupled to gravity[6].
We can redefine the fermion field in such a way that the kinetic
term decouples from the conformal mode at $D=2$.
The fermion mass term with the canonical dimension $2\Delta_0=1$
receives the gravitational dressing of the following form in terms
of the rescaled field
$$
\int d^D x \sqrt{g}^{1\over2} \bar \Psi \Psi
\ee
$$
In our gauge the coefficient of the kinetic term of the rescaled
fermion field is automatically fixed to be unity since the conformal
mode decouples at $D=2$.
Therefore we only need to consider the renormalization of the
operator $\sqrt{g}^{(1-\Delta_0)}$.

%%%
We now compare our result (4.10) with the exact solution of two
dimensional gravity [9 - 11]. In their conformal gauge approach, the
cosmological term $\int d^2 x \sqrt{\hat g}$ $ e^{-\phi}$ receives a
dressing from
gravitational fluctuations and becomes
$\int d^2 x \sqrt{\hat g} ~ e^{\alpha\phi}$.
Similarly the operator
$ \int d^2 x {({\sqrt{\hat g}}e^{-\phi})}^{1 - {\Delta_0}} \Phi$,
where
$\Phi$ is a general spinless primary field with scaling dimension
$2\Delta_0$, becomes $\int d^2 x {\sqrt{\hat g}}^{1 - {\Delta_0}}
e^{\beta \phi} \Phi$. The parameters $\alpha$ and $\beta$ have been
exactly computed :
$$
\eqalign{\alpha & = -{Q\over 2} \lbrace {
1-\sqrt{1-{8\over {Q^2}}} } \rbrace \cr
\beta & = -{Q\over 2} \lbrace {
1-\sqrt{1- {{8(1-\Delta_0)}\over Q^2}}
} \rbrace ~.  \cr} \ee
$$
In order to compare our result (4.10), we make use of the following
relation
$$
{\beta\over \alpha} = {{2(1-\Delta_0)+\gamma_{\Delta_0}}\over
{2+\gamma_{\Delta_0 = 0}}}  \ee
$$
which denotes the scaling dimension of the operator
${\int d^2 x (\sqrt{{\hat g}}e^{-\phi})}^{1 - {\Delta_0}} \Phi$
when we choose
as the standard scale that of the cosmological term.

This relation can be derived with recourse to the scaling argument
together with the renormalization group analysis. Let us consider
the following correlation functions
$$
 <\Pi_i \int d^D x{(\sqrt{\hat g} ~ e^{-{D\over 2}\phi})}^
{1-{\Delta_i}} \Phi_i> |_\mu ~.  \ee
$$
We rescale the background metric as ${\hat g}_{\mu\nu} \rightarrow
\lambda {\hat g}_{\mu\nu}$. The correlation function changes
$$
\Pi_i \lambda^{(1 - \Delta_i)} < \Pi_j \int d^D x
{(\sqrt{\hat g} e^{- {D\over2} \phi})}^{1 - \Delta_j} \Phi_j > |_
{{\mu\lambda}^{1\over 2}}  \ee
$$
where the renormalization scale also changes as $\mu \rightarrow
\mu \lambda^{1\over2}$. The renormalization group predicts
$$
(4.15) = \Pi_i \lambda^{(1-{\Delta_i})+{{\gamma_i} \over 2}}
< \Pi_j \int d^D x {(\sqrt{\hat g} e^{-{D\over 2} \phi})}^{1-\Delta_j}
\Phi_j> |_\mu  ~. \ee
$$
The running of the coupling constant $G$ can be neglected as
long as $G\gg \epsilon$.
If we choose the cosmological term as the standard scale, we obtain
(4.13).

Let us first evaluate (4.13) for the exact solution (4.12). By
expanding in powers of ${1\over{Q^2}}$, the result is
$$
{\beta\over \alpha} = {
       {
      1 - \Delta_0 + {1\over 2} {4\over {Q^2} }
   {(1- {\Delta_0} ) }^2
 + {1\over 2}{( {4\over {Q^2} }) }^2 {(1-{\Delta_0})}^3
+ O({ ( {4\over {Q^2} }) }^3)
         } \over
 {
  1 + {1\over 2} {4\over {Q^2}} + {1\over 2} {( {4\over {Q^2}})}^2
  + O({ ( {4\over {Q^2} }) }^3)
              }
                   } \ .  \ee
$$
On the other hand, if we use the perturbative values (4.9) and
(4.10) in (4.13), the result agrees with (4.17). Therefore our
computational scheme
is shown to give a ${1\over{Q^2}}={3\over{25-c}}$ expansion.

A comment is in order. We have assumed counter term dominance in
the one loop bare Lagrangian so $G_0$ is given by (4.8).
Therefore $G$ is assumed to be much larger than $\epsilon$.
However we may use the proposal of the present authors
(H.K. and M.N.[6]) to replace $G$ by the fixed point
of the $\beta$ function, $G^\ast = {{24\pi}\over{25-c}}\epsilon$.
We then obtain
$$ \gamma_{\Delta_0=0}(G^\ast)= -{4\over Q^2} - {4\over Q^2} +
{({4\over Q^2})}^2 + O ({({4\over Q^2})}^3) ~. \ee$$
Similarly
$$ \gamma_{\Delta_0} (G^\ast) = - {4\over Q^2}(1-\Delta_0) -
{4\over Q^2}(1-\Delta_0)+{({4\over Q^2})}^2 {(1-\Delta_0)}^3 ~.
\ee$$
This result disagrees with that of the exact solution (4.17).
Therefore, the conjecture that two dimensional gravity corresponds to
the $D=2$ limit of $D=2+\epsilon$ dimensional gravity at its fixed
points is regrettably incorrect. It rather appears to correspond to the
strong coupling phase of the $D=2+\epsilon$ dimensional quantum gravity.

Turning back to our two loop calculation, the following important
observation should be pointed out : The graphs containing the
$h_{\mu\nu}$, such as (c),(e) and (f), do not play any role in the
renormalization and thus we are allowed to consider only the
conformal mode $\phi$. This is what occurs in two dimensional
gravity in conformal gauge. Our observation would be further
confirmed by deriving the exact solution (4.12) in our scheme.
In fact in $2+\epsilon$ dimensions, we can derive (4.12) by
considering only the $\phi$ mode with the
assumption that the one loop counter term dominates in the bare
Lagrangian. By dropping the $h_{\mu\nu}$ field in Einstein
action (2.4),
we keep the following action for the $\phi$ field
$$
\sqrt{g} R \simeq \sqrt{\hat g} {\hat R}
e^{-{\epsilon\over 2}\phi}
-{{\epsilon(D-1)}\over 4} \sqrt{\hat g} ~
e^{-{\epsilon\over 2} \phi}
{\hat g}^{\mu\nu} \partial_\mu \phi \partial_\nu \phi ~.
\ee
$$
For later convenience a new variable $\psi$ is introduced through
$$
e^{-{\epsilon \over 4}\phi} = 1 + {\epsilon \over 4} \psi
\ee
$$
such that
$$
 \sqrt{g} R \simeq \sqrt{\hat g} {\hat R} (1 + {\epsilon \over 4}
\psi)^2-
{{\epsilon(D-1)}\over 4} \sqrt{\hat g} ~ {\hat g}^{\mu\nu}
\partial_\mu \psi \partial_\nu \psi  ~.
\ee
$$
Here, as in the perturbative calculation, we assume that the
following counter term (3.4) is dominant :
$$
 \eqalign{ {\cal L}_{c.t.} & = - {{25-c}\over{24\pi}}
{{\mu^\epsilon}\over \epsilon} \sqrt{\hat g} {\hat R}
{(1 + {\epsilon \over 4}\psi)}^2 \cr
& + {{25-c}\over{24\pi}} {{D-1}\over 4} \mu^\epsilon \sqrt{\hat g}
{\hat g}^{\mu\nu} \partial_\mu \psi
\partial_\nu \psi  ~ . \cr}
\ee
$$
This is free field theory for which the $\psi$ propagator is
$$
<\psi(P) \psi(-P)> = {{24\pi}\over{25-c}} {2\over{D-1}}
{1\over P^2} \quad .
\ee
$$
Thus, in computing the expectation value of the composite operator
$$
<{\sqrt{g}}^{1-{\Delta_0}}> = \sqrt{\hat g}^{1-\Delta_0}
< \hbox{exp} ~ \lbrace
{4\over \epsilon} (1-{\Delta_0}) \hbox{log}
(1 + {\epsilon \over 4}\psi) \rbrace >  \ee
$$
for a general operator $\int d^D x {(\sqrt{g})}^{1 - {\Delta_0}}
\Phi $, Wick's contraction theorem can be applied.
With each contraction, we associate a factor coming from the
divergent part of the $\psi$ loop
$$
{{24\pi}\over{25-c}} {2\over{D-1}} \int
{{d^D P}\over{{(2\pi)}^D}} {1\over P^2} =
- {24\over{25-c}} {1\over \epsilon} + 0(\epsilon^0) ~. \ee
$$

In order to evaluate exactly the divergent part of (4.25), we may
consider a zero dimensional model for which the action is given by
$$ S = {1\over 2} {{c-25}\over 24} \epsilon {\psi}^2  ~.
\ee$$
So eq.(4.25) reduces to the ordinary integration
$$
<{\sqrt{g}}^{1-\Delta_0}> = {1\over Z} \int^\infty_{-\infty}
d \psi ~ \hbox{exp} \lbrace
{4\over \epsilon}(1 - \Delta_0) \hbox{log}
(1+{\epsilon \over 4}\psi)
- {1\over 2} {{c-25}\over 24} \epsilon \psi^2 \rbrace
\ee
$$
where
$$
Z = \int^\infty_{-\infty} d \psi ~ \hbox{exp} (-{1\over 2}
{{c-25}\over 24} \epsilon \psi^2) ~. \ee
$$
By introducing another new variable
$\rho = {\epsilon\over 4 }\psi$, (4.28) reads
$$
<{\sqrt{g}}^{1-\Delta_0}> = \hbox{const.} \int^\infty_{-\infty}
d \rho  ~ \hbox{exp} {1\over \epsilon} \lbrace
4(1 - \Delta_0) \hbox{log} (1 + \rho) - {{c-25}\over 3} \rho^2
\rbrace ~.  \ee
$$

Let us evaluate (4.30) by means of the saddle point method.
The saddle point $\rho_0$ is given by
$$
\rho_0 = {1\over 2}
\{-1 \pm \sqrt{ 1- {{8(1 - \Delta_0)}\over{Q^2}}}\}
\ee
$$
which satisfies the equation
$$
{\partial \over{\partial \rho_0}} \lbrace
4(1 - \Delta_0) \hbox{log} (1 + \rho_0) -
{{c-25}\over 3} {\rho_0}^2 \rbrace = 0  ~ . \ee
$$
Evaluating (4.30) at $\rho=\rho_0$, we obtain the
renormalization of the
${\sqrt{g}}^{1-\Delta_0}$ operator as
$Z_{\Delta_0} {\sqrt{g}}^{1-\Delta_0}$ where
$$
Z_{\Delta_0} = \hbox{exp} \lbrace
-{4\over \epsilon} (1 - \Delta_0) \log (1 + \rho_0)
+ {{8\pi}\over {G_0}} \mu^{-\epsilon} {\rho_0}^2 \rbrace ~.
\ee
$$
Thus the anomalous dimension is given by
$$
 \gamma_{\Delta_0} = \mu {\partial\over{\partial \mu}}
  \log Z_{\Delta_0} = {\rho_0}^2 Q^2 = -2(1 - \Delta_0) - Q^2
\rho_0 ~.  \ee
$$
Inserting (4.34) into (4.13) we obtain the exact solution provided
that in (4.31) $\rho_0 = {1\over2}\lbrace{
-1 + \sqrt{1-{{8(1-\Delta_0)}\over{Q^2}}}
}\rbrace$ is chosen.\vfill\eject

\chapter{Conclusions and Discussions}

In this concluding section, we would like to clarify the basic
theoretical structure and physical picture of the quantum gravity
in $2+\epsilon$ dimensions.

Since the quantum gravity is not renormalizable in 4 dimensions,
we have pursued the $2+\epsilon$ dimensional expansion of quantum
gravity which is power counting renormalizable.

However the presence of the ${1\over\epsilon}$ pole in the propagator
of the conformal mode $\phi$ makes the renormalization program
difficult.
The problem is not the existence of the ${1\over\epsilon}$ pole
itself, since it is possible to choose a gauge as we have done in
which the interactions of $\phi$ are suppressed by powers of
$\epsilon$.

The real problem is how to handle the oversubtractions of $\phi$.
As we have explained before, the one loop counter term is an
oversubtraction for $\phi$.
If we consider insertions of the one loop counter term for $\phi$
$n$ times, it causes the extra singularities of
$O(({G \over \epsilon})^n)$.
This singularity cannot be subtracted by local counter terms in
general since the integrand itself is divergent before the loop
momenta integrations.
Hence it cannot be made finite by  differentiating the external momenta.
However these singularities do not originate from the high momentum
loop integrations and should not be regarded as the ultraviolet
singularities.
Therefore we have said that this problem is resolved by the
resummation of the one loop counter term insertions for $\phi$.
Putting it in another way we claim that the real expansion parameter
of the theory is not only $G$ but also $\kappa$ where
${\epsilon\over\kappa}$ is the effective inverse propagator of
the conformal mode.
We recall our bare Lagrangian
$$
{1 \over G_0} \sqrt{g} {R}{+}~ {\rm gauge~fixing~term}.
\ee
$$
The one loop quantum correction is
$$
{\mu^\epsilon \over \epsilon} {25-c \over 24\pi}
\sqrt{\hat g} \tilde{R} + {\rm finite~ terms}.
\ee
$$
By choosing ${1 \over G_0 }= \mu^\epsilon ({1\over G} -
{25-c \over 24\pi} {1 \over \epsilon})$, the theory becomes finite
up to the one loop level.
We stress that the only $h_{\mu\nu}$ field possesses the one loop
divergence.
For $\phi$, the \lq\lq tree level" Lagrangian is
$$
\mu^\epsilon({25-c \over 24\pi} - {\epsilon \over G})
({1\over4}\sqrt{\hat g}\tilde g^{\mu\nu}\partial_\mu\phi
\partial_\nu\phi
+ {1\over2}\phi\tilde R) + \cdots
\ee
$$
and the one loop correction is $O(\epsilon)$.
In other word at the one loop level $\kappa$ is nothing but
$G_0\mu^\epsilon$
$$
{1\over\kappa}=-{1\over G} + {25-c\over 24\pi\epsilon}.
\ee
$$
In this sense, the $h_{\mu\nu}$ field and $\phi$ field are
renormalized in very different ways.

Based on these understandings of the theoretical structure, we
present our basic physical picture of the quantum gravity in
$2+\epsilon$ dimensions.
The dynamics of $h_{\mu\nu}$ field resembles the spin system in
$2+\epsilon$ dimensions.
When the gravitational coupling $G$ is weak, the quantum gravity
is in the ordered phase.
When $G$ is strong it is in the disordered phase.
The ultraviolet fixed point is located at $G_c=O(\epsilon)$.
In the weak coupling regime, $G$ vanishes in the infrared limit.
The opposite situation takes place in the strong coupling regime.
When $G<G_c$, the theory contains massless gravitons if we can
extrapolate up to $\epsilon=2$.
It is physically the most interesting phase.
On the other hand the dynamics of the conformal mode is governed
by $\kappa$. $\kappa$ is of $O(\epsilon)$ away from the ultraviolet
fixed point, but it becomes larger near that point.
Therefore the conformal mode has nontrivial dynamics near the
ultraviolet fixed point.

Let us make the conformal mode $\phi$ massive by adding the
cosmological constant term $\Lambda \int d^Dx\sqrt g$ to the action.
If we do so, $\Lambda$ acts as the infrared cutoff like the
electron mass of QED and it sets the scale of the physics
(and the Universe).
By rescaling the metric $g_{\mu\nu}$, we fix $\Lambda=1$ further.
Then the bare coupling becomes $G_0\Lambda^{\epsilon\over D}$.
This is a free parameter of our theory which controls the size
of the Universe.

In the infrared limit $\Lambda\rightarrow0$, the effective bare
coupling $G_0\Lambda^{\epsilon\over D}$ vanishes.
In the previous section, the anomalous dimensions are computed in
power series of $G_0\mu^\epsilon\sim G_0\Lambda^{\epsilon\over D}$.
In the weak coupling limit, $h_{\mu\nu}$ field also become classical
in the infrared limit.
Therefore all anomalous dimensions due to gravitational dressing
vanish when $\Lambda\rightarrow0$ in the weak coupling regime.
The physics there is described in terms of classical Einstein theory
with small cosmological constant $G_0\Lambda^{\epsilon\over D}$.
This regime indeed resembles our Universe.
Since $G_0\Lambda^{\epsilon\over D}$ is a free parameter in
$2+\epsilon$ dimensional quantum gravity, we can tune it to have a
large Universe like our own.
On the other hand in the ultraviolet limit formally
$G_0\Lambda^{\epsilon\over D}\rightarrow\infty$ which means $\kappa$
becomes larger in that limit.
Let us introduce the effective central charge $c_{\rm eff}$ in such
a way that
$$
 {\epsilon\over\kappa}={25-c_{\rm eff}\over24\pi}.
\ee
$$
In terms of $c_{\rm eff}$, the ultraviolet limit corresponds to
$c_{\rm eff}=25$ limit.
In this sense, when $\epsilon$ is small enough the dynamics of the
conformal mode at the ultraviolet fixed point is very close to that
of the critical string and hence calculable.
The $O(\epsilon)$ corrections are calculable in the sense of
${1\over c}$ expansion[12].

The 2d gravity is scale invariant since all correlation functions
scale if we change the cosmological constant which controls the size
of the Universe.
In $2+\epsilon$ dimensions, the scale invariance is broken since the
dynamics of $h_{\mu\nu}$ and $\phi$ fields change in a nontrivial way
if we change the cosmological constant.

If we would like to have a constructive definition of the quantum
gravity in 3 or 4 dimensions, the theory should have ultraviolet
fixed points.
This condition puts the constraints in the matter content of the
theory such that $c<25$.
The number 25 is comfortably large.
It makes the applicability of $2+\epsilon$ dimensional expansion
of quantum gravity up to 4 dimensions plausible.
For such an enterprise, we make the following predictions:

The theory possesses two distinct phases, namely weak and strong
coupling phases.
We can control the theory by tuning the inverse of the bare
gravitational coupling constant
${1\over G_0}\Lambda^{-{\epsilon\over D}}$.

The continuum limit which resembles our Universe can be taken by
approaching the ultraviolet fixed point from the weak coupling side.

In the strong coupling phase near the phase transition point,
$c_{\rm eff}$ is in the range $1<c_{\rm eff}<25$.
Therefore the system may be in the branched polymer phase.

In fact these predictions are in accord with recent numerical
simulations[13].

Finally $c<25$ constraint may put the upper bound to the
quark-lepton species if we extrapolate $\epsilon$ up to 2.
Therefore the quantum gravity may explain why there are only three
generations of quarks and leptons in the Universe.

\smallskip
\centerline{\fourteenbf Acknowledgements}
\smallskip

%\ack
We are grateful to D. Lancaster for his reading the manuscript.
This work is supported in part by the Grant-in-Aid for Scientific
Research from the Ministry of Education, Science and Culture.

\smallskip
\noindent
{\fourteenbf Appendix ~Background field method in terms of $\phi$
and $h_{\mu\nu}$}

Let us consider the generating function of the connected Green's
functions in a background gauge,
$$
e^{-W_{\rm B.G.}}= \int \exp[ -S(g)-{\rm G.F.}(\hat g, h')+
J^{\mu\nu}\cdot h'_{\mu\nu} ],
\eqno{({\rm A}.1)}
$$
where $S$ is the Einstein action, G.F. denotes the gauge fixing
term which depends on the background $\hat g$ and $h'_{\mu\nu}$
field is defined as $g_{\mu\nu}=\hat g_{\mu\nu}+h'_{\mu\nu}$.
$\int$ implies the functional integration over the fields in the
theory including the ghost determinant.
$J^{\mu\nu}\cdot h'_{\mu\nu}$ implies $\int d^D x J^{\mu\nu}(x)
h'_{\mu\nu}(x)$.
Since our gauge fixing term is a simple function of $h_{\mu\nu}$
and $\phi$ fields in addition to $\hat g_{\mu\nu}$, it is certainly
a function of $h'_{\mu\nu}$ and $\hat g_{\mu\nu}$.
The effective action $\Gamma$ is obtained after the Legendre
transform
$$
\eqalign{
\Gamma &=W_{\rm B.G.} +J^{\mu\nu} < h'_{\mu\nu} > \cr
<h'_{\mu\nu}> &=- {\delta W_{\rm B.G.}\over \delta J^{\mu\nu}}
}\eqno{({\rm A}.2)}
$$
We may shift the quantum field such that
$h'_{\mu\nu}\rightarrow g_{\mu\nu}- \hat g_{\mu\nu}$.
Then
$$
e^{-W_{\rm B.G.}}= \int \exp[ -S(g)-{\rm G.F.}
(\hat g, g- \hat g)+ J^{\mu\nu}\cdot
(g-\hat g)_{\mu\nu} ]
$$
$$
\eqalign{
\Gamma &=W_{\rm B.G.} +J^{\mu\nu}\cdot
(<g_{\mu\nu}>-\hat g_{\mu\nu})\cr
       &= W+J^{\mu\nu}\cdot <g_{\mu\nu}>
}\eqno{({\rm A}.3)}
$$
where $W$ is the conventional generating function with an
unconventional gauge fixing term G.F.$(\hat g, g-\hat g)$.
In this way one can see that the background field method is an
efficient way to compute the conventional effective action in an
unconventional gauge.
By coupling the source term, we can expand the action around the
nontrivial background
$$
\eqalign{
S & (g) + {\rm G.F.} -J^{\mu\nu}\cdot g_{\mu\nu}\cr
= & S(\hat g) +
({\delta S\over \delta g}- J)^{\mu\nu}\cdot h'_{\mu\nu}
+ {\delta^2 S \over \delta^2 g}\cdot h'^2 + \cdots \cr
  & + {\rm G.F.} - J^{\mu\nu}\cdot \hat g_{\mu\nu}
}\eqno{({\rm A}.4)}
$$

When we compute the effective action in the background field
method, the linear term in $h'_{\mu\nu}$ field in the action
is dropped.
The logic behind it is the following relation
$$
J^{\mu\nu} = {\delta \Gamma \over \delta g^{\mu\nu}} =
{\delta S \over \delta g^{\mu\nu}} +
{\rm higher~order~terms~in~\hbar}
\eqno{({\rm A}.5)}
$$
Therefore if we use the conventional coupling of the source to
the metric $g_{\mu\nu}\cdot J^{\mu\nu}$, we should drop the linear
term in $h'_{\mu\nu}$.
In this paper we have adopted the parametrization of the metric
$g_{\mu\nu}=(\hat g e^h)_{\mu\nu}e^{-\phi}$, hence
$$
h'_{\mu\nu}= h_{\mu\nu} - \phi \hat g_{\mu\nu} + {1\over2}
(h^2)_{\mu\nu} -h_{\mu\nu}\phi +{1\over2} \phi^2\hat g_{\mu\nu}
+ \cdots
\eqno{({\rm A}.6)}
$$

If we drop only the linear term in $h_{\mu\nu}$ and $\phi$ in
the action, we have to add the following term in the action
$$
\eqalign{
&-{\delta S \over \delta g_{\mu\nu} }\cdot
\left( {1\over2} (h^2)_{\mu\nu} - h_{\mu\nu} \phi
+{1\over2}\phi^2\hat g_{\mu\nu} +\cdots \right)\cr
&= -\sqrt{\hat g}
\left({1\over2}\hat R\hat g^{\mu\nu} -\hat R^{\mu\nu}\right) \cdot
\left({1\over2}(h^2)_{\mu\nu}-h_{\mu\nu}\phi
+ {1\over2}\phi^2\hat g_{\mu\nu} + \cdots\right)
}\eqno{({\rm A}.7)}
$$
This additional term leads to the one loop divergence due to the
conformal mode
$$
-\sqrt{\hat g} {\epsilon\over 2} \hat R \cdot <\phi^2>
\eqno{({\rm A}.8)}
$$
However we also have the contribution to $\Gamma$ from the
source term
$$
\eqalign{
&J^{\mu\nu}\cdot<h'_{\mu\nu}>\cr
&= {\delta S \over \delta g_{\mu\nu} }\cdot
< {1\over2} (h^2)_{\mu\nu} - h_{\mu\nu} \phi
+{1\over2}\phi^2\hat g_{\mu\nu} +\cdots >\cr
&= \sqrt{\hat g}{\epsilon\over 2}\hat R \cdot <\phi^2>
}\eqno{({\rm A}.9)}
$$
These two divergences cancel each other.
In conclusion, we can simply drop the linear term in $h_{\mu\nu}$
and $\phi$ in the action to compute the one loop divergence in
the effective action.

\endpage

{\bf Figure caption}
\hskip 40pt \itemitem{Fig.}1 The two loop correction graphs
to $\sqrt{{\hat g}}$ which is shown by a cross.
The solid and wavy lines denote $\phi$ and
$h_{\mu \nu}$ propargators, while the dashed and
dash-and-dotted lines are $\psi^\mu$ and matter propagators.
\vskip 40pt
{\bf References}

\item{[1]}M.E. Agishtein and A.A. Migdal, PUPT-1287~(1991).
\item{[2]}J. Ambj\o rn and S. Versted, NBI-HE-91-45~(1991).
\item{[3]}S. Weinberg, in General Relativity, an Einstein
Centenary Survay, eds. S.W. Hawking and W. Israel
(Cambridge University Press, 1979) p.790.
\item{[4]}R. Gastmans, R. Kallosh and C. Truffin, Nucl. Phys.
\undertext{B133}~(1978)~417.
\item{[5]}S.M. Christensen and M.J. Duff, Phys. Lett.
\undertext{B79} (1978) 213.
\item{[6]}H. Kawai and M. Ninomiya, Nucl. Phys.
\undertext{B336}~(1990)~115.
\item{[7]}G. 't Hooft and M. Veltman, Ann. Isnt. Henri Poincare
\undertext{20}~(1974)~69.
\item{[8]}L.F. Abbott, Nucl. Phys. \undertext{B185} (1980) 189.
\item{[9]}J. Distler and H. Kawai, Nucl. Phys.
\undertext{B321}~ (1989)~509.
\item{[10]}F. David, Mod. Phys. Lett.~\undertext{A3}~ (1988)~1651.
\item{[11]}V.G. Knizhnik, A.M. Polyakov and A.B. Zamolodchikov,
Mod. Phys. Lett. \undertext{A3}~(1988)~819.
\item{[12]}H. Kawai, Y. Kitazawa and M. Ninomiya, paper
in preparation.
\item{[13]}M.E. Agishtein and A.A. Migdal, PUPT-1311 (1992).

\end